\documentclass[preprint,amsmath,amssymb,prb]{revtex4-2}

\usepackage[utf8]{inputenc}
\usepackage{natbib}
\usepackage{graphicx}

\usepackage{amssymb}

\usepackage{lineno}

\usepackage[normalem]{ulem}
\usepackage{tikz}
\usetikzlibrary{positioning,shapes,shadows,arrows}
\usepackage{comment}
\usepackage{mathtools}
\usepackage[T1]{fontenc}

\begin{document}

\preprint{PRB}



\title{A topologically-derived dislocation theory for twist and stretch moir\'e superlattices in bilayer graphene}




\author{Emil Annevelink}
\affiliation{
 Department of Mechanical Science and Engineering, University of Illinois at  Urbana-Champaign, Urbana IL 61801 USA 
}%

\author{Harley Johnson}
\affiliation{
 Department of Mechanical Science and Engineering, University of Illinois at  Urbana-Champaign, Urbana IL 61801 USA \\
 Department of Materials Science and Engineering, University of Illinois at  Urbana-Champaign, Urbana IL 61801 USA
}%

\author{Elif Ertekin}
\affiliation{
 Department of Mechanical Science and Engineering, University of Illinois at  Urbana-Champaign, Urbana IL 61801 USA \\
}%


\begin{abstract}

We develop a continuum dislocation description of twist and stretch moir\'e superlattices in 2D material bilayers. 
The continuum formulation is based on the topological constraints introduced by the periodic dislocation network associated with the moir\'e structure. 
The approach is based on solving analytically for the structural distortion and displacement fields that satisfy the topological constraints, and which minimize the total energy. 
The total energy is described by both the strain energy of each individual distorted layer, and a Peierls-Nabarro like interfacial contribution arising from stacking disregistry.
The dislocation core emerges naturally within the formalism as a result of the competition between the two contributions. 
The approach presented here captures the structure and energetics of twist and stretch moir\'e superlattices of dislocations with arbitrary direction and character, without assuming an analytical solution {\it a priori}, with no adjustable parameters, while accounting naturally for dislocation-dislocation image interactions. 
In comparisons to atomistic simulations using classical potentials, the maximum structure deviation is 6\%, while the maximum line energy deviation is 0.019 eV/\AA.
Several applications of our model are shown, including predicting the variation of structure with twist angle, and describing dislocation line tension and junction energies.
\end{abstract}

\maketitle

\section{Introduction}

Moir\'e superlattices are periodic patterns created when two lattices are stretched or rotated with respect to one another \cite{Hermann_2012}.
The stretch or rotation gives rise to unique electronic properties distinct from the undistorted system. 
For instance, the moir\'e patterns that form from two layers of two-dimensional materials such as bilayer graphene create a unique platform for studying exotic effects such as superconductivity and correlated electron physics \cite{Wong2020,Uri2020,Liu2020,Cao2018}.

\begin{figure*}[t]
\centering
\includegraphics[width=\linewidth]{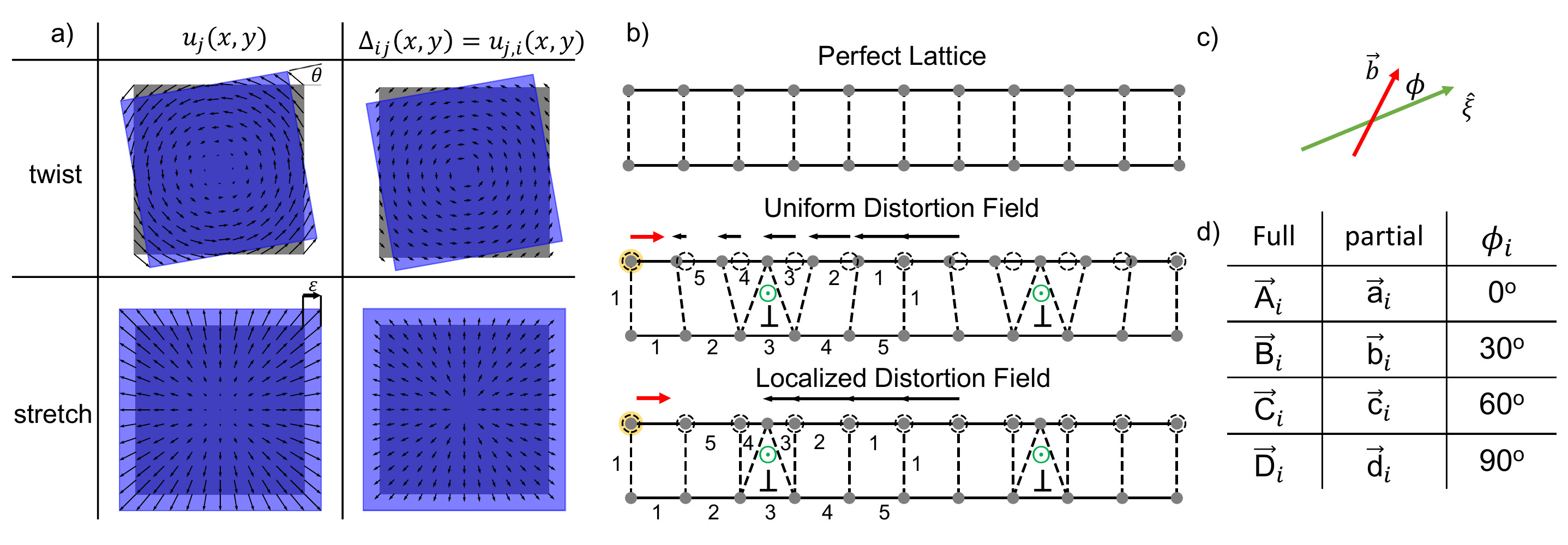}
\caption{(a) Displacement $u_j$ and distortion $\Delta_{ij}$ fields of a uniform twist or stretch moir\'e superlattice, giving linear displacement and constant distortion fields. 
The fields are related to each other by a spatial derivative given in Einstein notation.
(b) The displacement fields (black arrows) operate on atomic positions from a perfect lattice.
A Burgers circuit, where numbers around the dislocation count lattice sites, reveals the identical topological characteristic, the Burgers vector (red), of the dislocations both uniform and localized distortion.
(c) The topological components that define a dislocation are the Burgers vector $\vec{b}$ (red), the line direction $\hat{\xi}$ (green), and the sense $\phi$.
(d) Catalog of dislocations. Full and partials are given by upper and lower case letters respectively, where A-D are the primary types in a triangular lattice.}
\label{fig:intro_combined}
\end{figure*}

In a moir\'e superlattice, displacement $u_j$ and distortion $\Delta_{ij} = \partial_i u_j = u_{j,i}$ fields define the relative shift between the two layers measured from a reference.
For example, pure twist and stretch moir\'e patterns have displacement fields that vary linearly with distance from the origin and constant distortion tensor components. 
Figure \ref{fig:intro_combined}(a) shows examples of both.  
However, pure twists or stretches in real materials are rare. 
Local internal relaxations, if permitted, may shift atomic positions from the idealized fields shown in Figure \ref{fig:intro_combined}(a) to minimize the configuration energy. 
Thus, rather than pure twists or stretches, distorted regions tend to become localized and separated from each other by large regions that are almost entirely undistorted.
At the atomic scale, the localization of the deformed region increases regions of stacking registry and reduces regions of disregistry. 
The rearrangement into regions of large and small distortion corresponds to the formation of interlayer dislocations. 
In Figure \ref{fig:intro_combined}(b), ideal uniform and localized distortions of a mock 1D bilayer system for a stretch moir\'e are illustrated. 
In the former, the disregistry is uniform while in the latter it is localized to well-defined regions corresponding to the location of an edge dislocation. 
Topologically, however, the uniform and localized cases are identical.

The shared topological feature is a stacking fault that separates distinct regions of lattice stacking.
In twisted bilayer graphene moir\'e superlattices, stacking faults have been observed experimentally as regions that separate AB and AC (or BA) stacking.
The stacking fault has been described mathematically as a soliton and observed with dark-field transimission electron microscopy to analyze the width of its core \cite{Alden2013}.
From a topology perspective, the stacking fault is an interlayer dislocation. 
Using classical potentials, both Zhang {\it et al.}\ \cite{ZHANG2017,ZHANG2018} and Gargiulo and Yazyev \cite{Gargiulo_2017} identified the moir\'e wavelength of dislocation localization.
Continuum models have predicted that even finer scale modulations of the dislocation and moir\'e superlattice structure may be present \cite{Dai2016_Nano,Dai2016_PRB}.
Together, these set the foundation that dislocation descriptions can effectively describe the structure of moir\'e superlattices, as recently suggested by Gornostyrev and Katsnelson \cite{Gornostyrev_2020Arxiv}. 
However, in order to confidently use continuum dislocation descriptions of moir\'e superlattices, a formal treatment to establish the equivalence of interlayer dislocations and moir\'e superlattice topology is needed.

In this work we formalize a linear elastic theory of bilayer graphene interlayer dislocations, and rigorously link them to moir\'e superlattices.
Our approach is distinct as we account for the dislocation geometries explicitly through the topological constraints that they introduce in the displacement and distortion fields.
The solution is obtained by solving for the fields that minimize the total energy while satisfying the required topology. 
The structure of the dislocation core arises as a result of a competition between intra-layer strain energy and inter-layer interface energy.
Our approach naturally accounts for moir\'e superlattice periodicity, including dislocation -- dislocation interactions that can alter the core structure (such as for large twists or stretches).
The resulting formalism has no adjustable parameters (model parameters are found first, directly from interatomic potentials), and does not {\it a priori} assume an analytical form for the solution. 

Our approach correctly reproduces the energies and displacement fields obtained from atomic scale simulations using classical potentials.
To highlight applications of our method, we show how the dislocation core structure evolves with varying twist angle, which reveals the AA stacking that prevails at large twists to be a result of core interactions. 
We also estimate line and junction energies of arbitrary dislocations in bilayer graphene, and find that 0$^\circ$ dislocation junctions are attractive and 90$^\circ$ dislocation junctions are repulsive.

\begin{figure*}[h]
\centering
\includegraphics[width=6.45in]{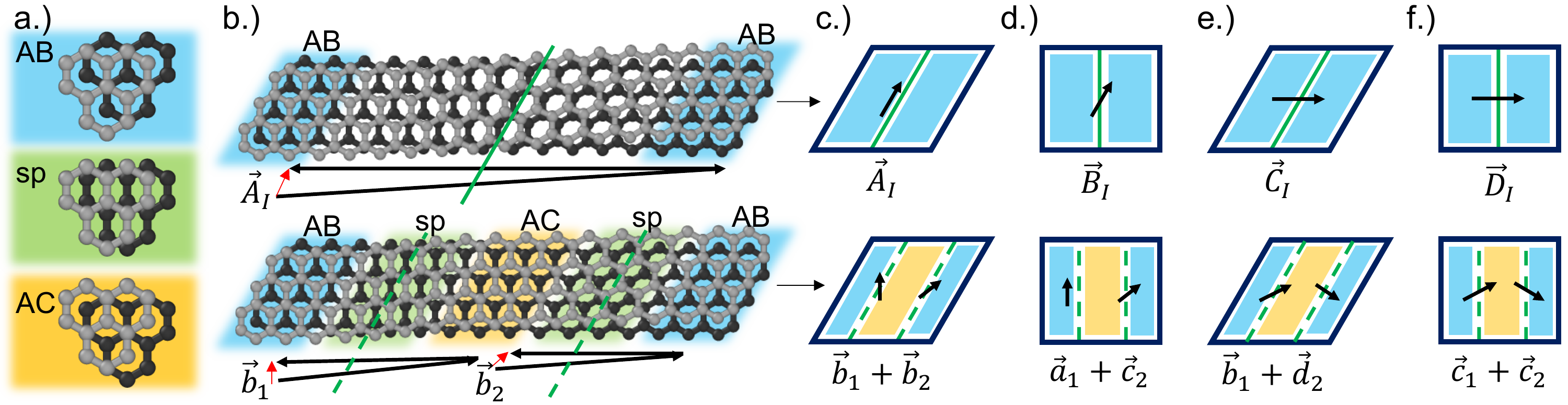}
\caption{Structure of dislocations in bilayer graphene.
(a) The possible stackings of bilayer graphene. AB/AC are degenerate low energy stacking, while sp stacking is the saddle point energy separating AB and AC regions. 
(b) Ball and stick representation of 0$^\circ$ full dislocation (top) that separates into two 30$^\circ$ partial dislocations (bottom) with associated Burgers circuits. 
The full dislocation Burgers circuit traverses 15 lattice vectors in the top and bottom layer yielding the closure failure shown in red along the solid green dislocation line. 
The two partial dislocations Burgers circuits traverse 7 lattice vectors in each layer yielding closure failures both 30$^\circ$ relative to the dotted green line.
(c) Continuum representation of full and partial dislocations from (b), showing a 0$^\circ$ full dislocation $\vec{A}_1$ and two 30$^\circ$ partial dislocations $\vec{b}_1$, $\vec{b}_2$.
(d-f) Three remaining full dislocation directions, respectively 30$^\circ$, 60$^\circ$, and 90$^\circ$  and their partials, respectively 0$^\circ$/60$^\circ$, 30$^\circ$/90$^\circ$, and 60$^\circ$/60$^\circ$.}
\label{fig:1ddislocations}
\end{figure*}

\section{Geometry of interlayer dislocations in bilayer graphene}

The presence or absence of a dislocation is determined from Burgers circuits formed around a region of material.
For example, in Figure \ref{fig:intro_combined}(b), a Burgers circuit with a right--handed, start--finish (RH--SF) convention \cite{Frank_1951} around both the linear and localized stretch moir\'e structure encloses a dislocation with line direction  coming out of the page (green). 
Starting at the top left, five steps are used to move along the layers and one step is used to traverse between them. 
The Burgers vector $b$ (red) is the closure failure of the loop and quantifies the incompatibility in the displacement fields. 
It is identical for the linear and the localized case and equal to the lattice vector. 
The presence of the edge dislocation is denoted by the symbol $\bot$. 
The two cases correspond respectively to an infinitely distributed or infinitely localized core.

The topological character of a dislocation is defined by Burgers vector $b$ and dislocation line $\xi$ (Figure \ref{fig:intro_combined}(c)). 
The dislocation line defines the direction, and the Burgers vector describes the magnitude and direction of the incompatibility in the displacement field.
The angle $\phi$ between $b$ and $\xi$ determines the sense of the dislocation (edge, screw, or mixed). 
In Figure \ref{fig:intro_combined}(b) $\phi = 90^\circ$, but in triangular lattices like bilayer graphene there are four crystallographic dislocations with unique angles. In Figure \ref{fig:intro_combined}(d), they are presented as letters, where full dislocations and partial dislocations are differentiated by their capitalization.

\subsection{One-dimensional dislocation networks}

Full dislocations (Figure \ref{fig:1ddislocations}(b)) are boundaries separating regions of AB stacking \cite{Butz2014} and so have Burgers vectors of magnitude equal to the lattice vector. 
The four crystallographic full dislocations (four total) are shown in the top row of Figure \ref{fig:1ddislocations}, through (b) atomistic and (c-f) continuum representations. 
Using a right-handed start-finish (RH-SF) Burgers circuit that traverses from AB stacking on the left to AB stacking on the right along the top gray layer and back along the bottom black layer, the closure failure yields the Burgers vector (A$_I$, red).
It is parallel to the dislocation line (green) and has a sense $\phi$=0$^\circ$.

Full dislocations are rarely observed in graphene bilayers since the two atom basis permits the splitting of dislocations into partials that separate regions of equivalent AB and AC stacking (Figure \ref{fig:1ddislocations}(a)).
Partial dislocations have a high-symmetry SP stacking halfway between the AB and AC stacking centered at the dislocation line, as shown in the bottom row of Figure \ref{fig:1ddislocations}(b).
The structure of the full and partial dislocations in Figure \ref{fig:1ddislocations}(b) differ by the relaxation to AC stacking in the central region of the latter.
The relaxation decomposes the full dislocation $A$ into two partials.
The two partials are labeled b$_1$ and b$_2$ according to their 30$^\circ$ sense.
The topological characteristic of the isolated full and two partials are the same, creating the dislocation reaction A$_I$=b$_1$+b$_2$.

\begin{figure*}[h!]
\centering
\includegraphics[width=6.45in]{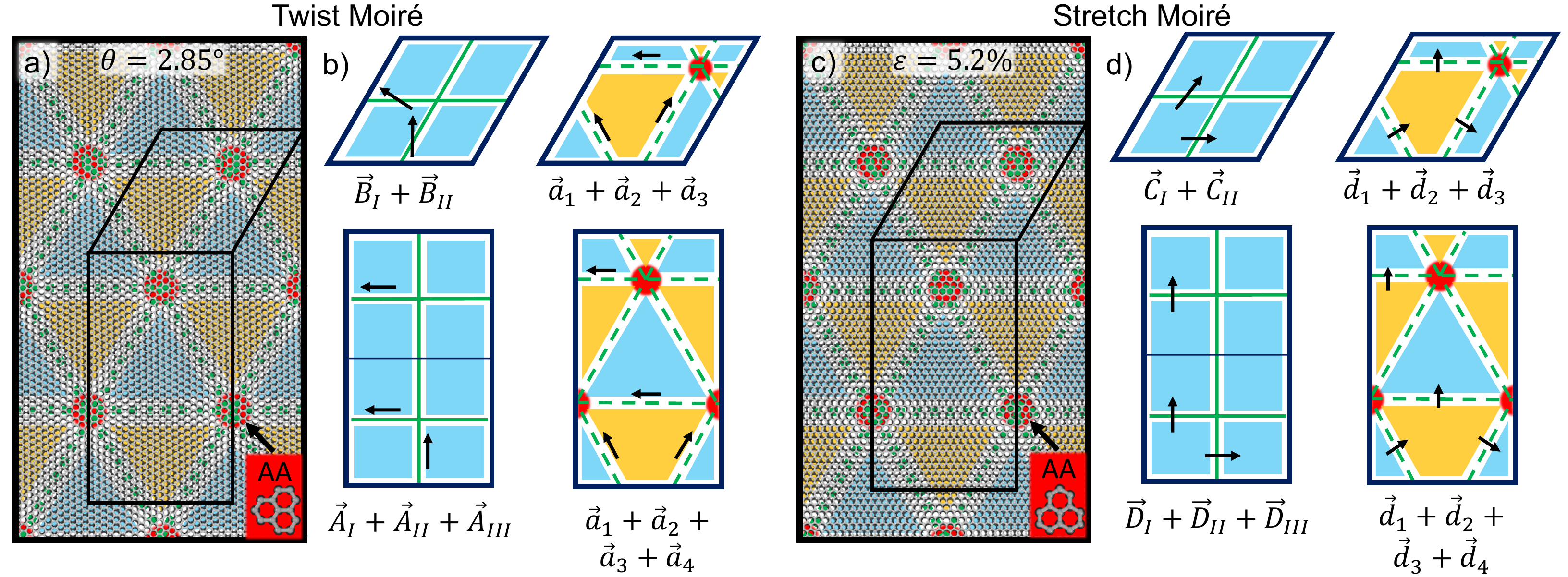}
\caption{Twist and stretch moir\'e patterns are equivalent to 2D networks of, respectively, 0$^\circ$ and 90$^\circ$ partial dislocations.
(a) Twisted bilayer graphene ($\theta=2.85^\circ$) resulting in a twist moir\'e pattern.
Triangular and rectangular supercells are overlaid to show possible periodic computational domains. 
The red circles are regions of high energy AA stacking (inset) that correspond to partial dislocation junctions.
(b) Continuum representations of a twist moir\'e pattern of full and partial dislocation configurations for triangular and rectangular supercells.
The twist moir\'e is a 2D network of partial dislocations parallel to their line direction ($\phi=0^\circ$).
(c) Bilayer graphene with one layer bi-axially stretched over the other ($\varepsilon=5.2\%$) resulting in a stretch moir\'e pattern.
(d) Continuum representation of a stretch moir\'e pattern in terms of full and partial dislocations for triangular and rectangular supercells.
The stretch moir\'e is a network of partial dislocations with Burgers vectors perpendicular to their line direction ($\phi=90^\circ$).
}
\label{fig:2Ddislocations}
\end{figure*}

\subsection{Moir\'e structures: two-dimensional dislocation networks}

Moir\'e superlattices are equivalent to  two-dimensional networks of dislocations \cite{POCHET2017}.
For bilayer graphene, we identify the dislocation networks for twist and stretch moir\'e superlattices. 
Compared to 1D networks, 2D networks may include junctions of dislocation lines that correspond to high energy AA stacking in bilayer graphene (inset Figure \ref{fig:2Ddislocations}(a,c)).

A ball-and-stick representation of perfect twist deformation of 2.85$^\circ$ is shown in Figure \ref{fig:2Ddislocations}(a).
Regions are shaded by the stacking type which reveals the moir\'e superlattice. 
The triangular symmetry is visible immediately.
Two possible supercells,  rectangular and triangular, are shown.
Using a Burgers circuit, the $\vec{B}_i$ dislocations in the triangular supercell split into three $\vec{a}_i$ dislocations \cite{POCHET2017}.  
Equivalently, using the rectangular supercell three $\vec{A}_i$ dislocations split into four $\vec{a}_i$ dislocations.
Therefore, a twist moir\'e superlattice corresponds to a periodic network of partial screw dislocations with dislocation lines oriented at 60$^\circ$ to each other.  
The twist angle determines the size of the superlattice and the dislocation spacing.

Similarly, stretch moire superlattices are described by triangular networks of partial dislocations but with a 90$^\circ$ edge sense.
The ball and stick representation in Figure \ref{fig:2Ddislocations}(c) shows a perfect stretch moir\'e with 5.2\% strain.
A key difference between Figure \ref{fig:2Ddislocations}(a,c) is a 90$^\circ$ rotation of the upper layer (visible in the AA insets).
So, although the dislocation line structure looks identical, the Burgers vectors are rotated by 90$^\circ$.
This gives the dislocation reactions for triangular unit cells of two $\vec{C}_i$ dislocations to three $\vec{d}_i$ dislocations or for rectangular unit cells three $\vec{D}_i$ dislocations to four $\vec{d}_i$ dislocations.

\section{Continuum model for interlayer dislocations}

The approach to describe interlayer dislocations in bilayer graphene is based on a continuum formalism of the structure and energy of periodic dislocation networks originally formulated by Mura \cite{Mura1964}, later adapted by Daw \cite{DAW2006}, and then applied to the description of topological defects in monolayer graphene \cite{Ertekin2009,Chen2011, ANNEVELINK2019}.
It is based on the idea that each dislocation introduces a {\it topological constraint} that must be satisfied by the distortion fields $\Delta_{ij}$.
The solution is obtained by finding the distortion that satisfies the topological constraints, while using any remaining degrees of freedom to minimize the total energy. 

The method developed here adapts the original formulation of Daw to the case of interlayer dislocations in bilayer graphene.
Compared to existing descriptions of interlayer dislocations in bilayer graphene \cite{Dai2016_PRB,Gornostyrev_2020Arxiv}, desirable features of our approach are (i) that solutions are  obtained directly without the need to assume an analytical form, (ii) dislocation -- dislocation interactions and periodic boundary conditions are naturally accounted for, and (iii) no model parameters are adjusted to fit to the atomistic results.

\subsection{Total and Elastic Energy}

The total energy of a deformed bilayer is  
\begin{equation}
E_{tot} = E^1_{elastic}+E^2_{elastic}+E_{interface} \hspace{0.5em},
\label{Eq:E_tot_short_flat}
\end{equation}
with an elastic term for each layer and an interface energy that couples the layers.
The interface energy contribution is discussed in Section \ref{sec:interface}.
The elastic energy for layer $I=1,2$ is given by the integral of the strain energy density, or
\begin{equation}
E^I_{elastic} = \frac{1}{2}C_{ijkl} \int_{cell} \Delta^I_{ij}\Delta_{kl}^{I*} dA  = \frac{\Omega_A}{2} \sum_GC_{ijkl}\widetilde{\Delta}^I_{ij}\widetilde{\Delta}^{I*}_{kl},
\label{Eq:E_elastic_flat_short}
\end{equation} 
where $C_{ijkl}$ are intra-layer elastic constants and $\Delta^I_{ij}$ is the distortion tensor for layer I. 
Einstein notation, where repeated indices are summed, is used.

By definition, the distortion field exhibits the periodicity of the moire superlattice and can be expressed as a Fourier series, or \begin{equation}
    \Delta_{ij}(X) = \sum_G \widetilde{\Delta}_{ij}(G) \, \exp(iG\cdot X) \hspace{0.5em}, \label{fs}
\end{equation}
where the summation is over reciprocal lattice vectors of the moire superlattice $G$, reciprocal components are distinguished using a tilde $\widetilde{\Delta}$, and the distortion tensor is a spatially varying field of position X.  
Substituting Equation (\ref{fs}) into the integral in Equation (\ref{Eq:E_elastic_flat_short}) gives the summation on the right hand side, where $\Omega_A$ is the area of the moir\'e superlattice unit cell.
We consider bilayers constrained to remain flat, which we will show results in a linear system of equations that can be directly solved for distortion tensor components $\widetilde{\Delta}^1_{ij},\widetilde{\Delta}^2_{ij}$ (Section \ref{sec:en_min}).

\subsection{Topological Constraints for Interlayer Dislocations}

In typical bulk materials, the presence of a dislocation is indicated by a topological constraint given by the Nye tensor
\begin{equation}
\alpha_{jm} = \epsilon_{jkl}\partial_k\Delta_{lm} = \xi_jb_m\delta(r_\bot) \hspace{0.5em},
\label{Nyetensor_cont}
\end{equation}
where $\xi_j$, $b_m$, and $r_\bot$ are respectively the dislocation line direction, Burgers vector, and the perpendicular distance to the dislocation line $\xi$ \cite{NYE1953} . 
The Nye tensor introduces an incompatibility into the displacement field wherever a dislocation is present, as indicated by the curl of the distortion tensor $\Delta_{lm}$.
Compared to bulk dislocations, the formulation for interlayer dislocations in 2D bilayers makes two sets of changes to Equation (\ref{Nyetensor_cont}).

The first set arises from the bilayer nature of 2D materials.
We treat the bilayer as two isolated 2D layers that are continuous in--plane, but coupled to each other in the third direction via interfacial energy $E_{interface}$ in Equation (\ref{Eq:E_tot_short_flat}). 
This causes the repeated indices in Equation (\ref{Nyetensor_cont}) to be summed over only the two in--plane directions while the continuous partial derivative $\partial_3$ in the out--of--plane direction is replaced by a discrete difference between the two layers. 
Additionally, for interlayer dislocations, the Burgers vector $b$ and line direction $\xi$ only have components in the two in-plane directions. 

The second change pertains to modifying the delta function in Equation (\ref{Nyetensor_cont}). 
In the original formulation, the presence of the delta function causes the elastic energy to diverge.
To remove the divergence, it is typically smoothed into a gaussian, and normalized so that the integrated total incompatibility is fixed to the magnitude of the Burgers vector. 
The smoothing causes the elastic energy to become finite, decreasing monotonically with the width of the gaussian.  
The width is referred to as the core radius, since it indicates the spatial extent of the dislocation core.
The narrow core limit (see `localized' in Figure \ref{fig:intro_combined}(b)) resembles the original delta function that causes an infinite elastic energy.
The infinite core limit, corresponding to a uniform distortion everywhere (see `uniform' in Figure \ref{fig:intro_combined}(b)), gives the minimum elastic energy. 
Typically the core radius is the only adjustable parameter, and is fitted to reproduce total energies as obtained from atomistic simulations. 

In our formulation, the interface contribution to the total energy $E_{interface}$ in Equation (\ref{Eq:E_tot_short_flat}) penalizes large core radii since they introduce extended regions of stacking disregistry.  
The interface energy, in contrast to the elastic energy, is largest with distributed core and smallest with a localized core (it is described in detail in the following subsection).   
The inclusion of the interface energy allows us to generalize the topological constraint and formulate it in terms of the average value of the  incompatibility inside the moir\'e superlattice. 

Accounting for these modifications, the topological constraint adopts the generalized form 

\begin{equation}
\Omega_A\epsilon_{j3l}\langle \Delta^{1}_{lm}-\Delta^{2}_{lm} \rangle = \xi_jb_m\hspace{0.5em},
\label{Eq:Nyetensor_disc_integrated}
\end{equation}
where the finite difference between the distortion tensors in each layer comes from the first set of changes, while the average of the difference comes from the second set.
Rather than an explicit predefined core radius, an effective core radius emerges as a result of the competition between in--plane elastic energy and stacking energy.  
This results in a core structure that arises from the competition, and a model with no adjustable parameters that are fit to atomistic results.
Instead, all model parameters are fit to best represent the interatomic potentials.

\subsection{Interface Energy Contribution}
\label{sec:interface}

\begin{figure*}[h!]
\centering
\includegraphics[width=6.45in]{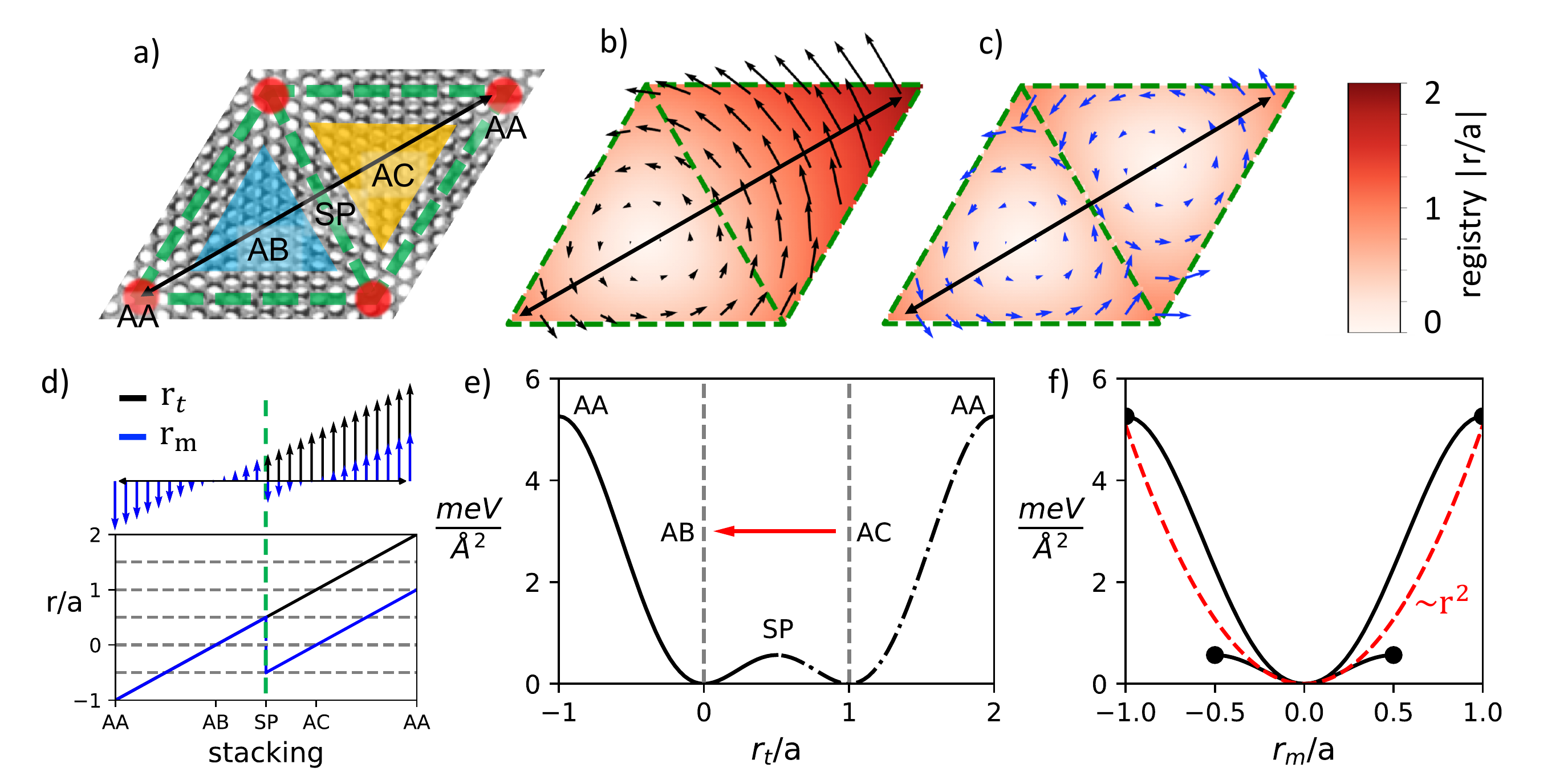}
\caption{(a-d) Local registry function r$_m$ and (e-f) harmonic stacking fault energy.
(a) Twisted bilayer graphene with dislocation lines and stacking regions.
(b) Full registry of one layer relative to the other centered at an AB stacking location. 
For a uniform twist, the magnitude of the registry increases linearly with distance from the AB stacking center.
(c) The folded registry describes the registry relative to the closest AB/AC stacking location, which always has a normalized magnitude less than one.
(d) Line traces of the twist and folded registry functions from (b) and (c).
(e) Bilayer graphene stacking fault energy for a rigid translation along the armchair direction (AA to AA) with a constant interlayer spacing of 3.4\AA, which shows the degenerate AB/AC minima as well as the energies of AA and SP.
(f) Approximate harmonic interface potential (red) is found by fitting the critical points (AA, SP, AB) of the shifted interface potential.}
\label{fig:linearinterfacemodel}
\end{figure*}

The interface energy accounts for the disregistry between the layers similar to the Peierls-Nabaro model \cite{Peierls_1940,Nabarro_1947}.
We restrict the interface energy to the same form as the elastic energy (summation over squares), but now the summation is over displacement differences between the layers. 
The interface energy is given by
\begin{equation}
E_{interface} =A_{jl} \int_{cell} r_{j}r_{l}^{*} dA = \Omega_A \sum_G
A_{jl}\widetilde{r}_j\widetilde{r}^{*}_l \hspace{0.5em},
\label{Eq:E_interfaceu_flat}
\end{equation}
where $A_{jl}$ is a proportionality constant analogous to the elastic constants in Equation (\ref{Eq:E_elastic_flat_short}), and $r$ is the local registry given by the difference of displacement fields of each layer $u^1-u^2$.  

As shown in Figure \ref{fig:linearinterfacemodel}(a-d), the expression for the interface energy is valid when $r = r_m$ (mapped registry, defined with respect to the closest minima) rather than for $r = r_t$ (total registry, defined from a single reference point).
Figure \ref{fig:linearinterfacemodel}(a) shows the stackings, while \ref{fig:linearinterfacemodel}(b,c) give the total and mapped registries for a perfect twist.
The total registry $r_t$ increases linearly with distance from a selected AB center and can have magnitudes greater than the carbon-carbon distance $a$.
The mapped registry $r_m$ has no value larger than $a$.
For mapped registry, AB/AC stacking both have $r_m=0$, SP stacking has $r_m/a = 0.5$, and AA stacking has the maximum registry of $r_m/a = 1$.
Line scans for the total and mapped registry fields are shown in Figure  \ref{fig:linearinterfacemodel}(d) to show that the mapped registry is obtained by subtracting the Burgers vector from the total registry when traversing a dislocation.

The generalized stacking fault energy for flat bilayer graphene is shown in Figure \ref{fig:linearinterfacemodel}(e).  
It is obtained using classical force fields by rigidly sliding one layer relative to the other at constant interlayer spacing along the armchair direction from AA to AA stacking. 
By fitting the parameters $A_{jl}$ from Equation \ref{Eq:E_interfaceu_flat} to the stacking fault energy, we obtain a harmonic description.
The fitted approximation of the stacking fault energy shown in Figure \ref{fig:linearinterfacemodel}(f) overestimates the SP and underestimates the AA energy, decreasing the relative size of the SP and increasing the relative size of the AA regions.
Forcing the curvature from AB to SP to be the same as that of AB to AA has the consequence that we cannot capture the finer scale change in concavity of the registry contours near the junction as reported by Gargiulo {\it et al.} using atomistic simulations.

\subsection{Energy Minimization}

\label{sec:en_min}

Using the expression for the total energy in Equation \ref{Eq:E_tot_short_flat}, we find distortion fields $\Delta_{ij}$ that minimize the total energy while satisfying the topological constraints in Equation \ref{Eq:Nyetensor_disc_integrated} imposed by the dislocations network. 
The minimum energy is found in reciprocal space; the Fourier transform of the topological constraint gives 
\begin{equation}
    \widetilde{\Delta}^{1}_{lm}(G=0)-\widetilde{\Delta}^{2}_{lm}(G=0) = \frac{1}{\Omega_A}\epsilon_{j3l}\xi_jb_m \hspace{0.5em}, 
\end{equation}
a contribution only for the $G = 0$ Fourier coefficients. 
When constrained to be flat, the minimum elastic energy is achieved by sharing the topological constraint equally between the layers ($\widetilde{\Delta}^1$=$-\widetilde{\Delta}^2$).
The interface energy does not affect the distribution of the topological constraint since changing the distribution does not change the registry between layers. 

The solution is separated into inhomogeneous and homogeneous components $\Delta = \Delta^{inh}+\Delta^{hom}$, where the former satisfies the constraints. 
The homogeneous term is the general solution that does not change the dislocation content ({\it e.g.}\, $\alpha=0$).
In reciprocal space, the homogeneous solution satisfies $G \times \widetilde{\Delta}^{hom} = 0$ and corresponds to displacement fields that are compatible.
This gives the general form 
\begin{equation}
    \widetilde{\Delta}^{I,hom}_{ij}=G_i \widetilde{\chi}^I_j \hspace{0.5em},
    \label{eq:homogeneous}
\end{equation}
where the vector $\widetilde{\chi}$ encompasses the remaining degrees of freedom in $\widetilde{\Delta}$. 
For each $G$, its two components are determined by minimizing the energy (Equation \ref{Eq:E_tot_short_flat}) with respect to them. 
Different $G$-components enter the energy separately in the sum, so this can be done algebraically by solving $\partial E_{tot}/\partial \widetilde{\chi}^{1*}_l=0$ and $\partial E_{tot}/\partial \widetilde{\chi}^{2*}_l=0$ simultaneously. 
Further details of solving the partial differential equations are found in the supplementary information.
The topological constraint in Equation \ref{Eq:Nyetensor_disc_integrated} introduces only non-zero $\widetilde{\Delta}^{inh}$ for $G = 0$. 
But the folded displacement field appears in the expression for the interfacial energy, which has the consequence of introducing non-zero $\widetilde{\Delta}^{hom}$ for all $G$.
The detailed solution is shown in the Supporting Information.

\section{Comparison to classical potential atomistic simulations}

We apply our dislocation formalism to the 1D and 2D dislocation networks shown in Figures \ref{fig:1ddislocations} and \ref{fig:2Ddislocations} and compare them to atomic scale simulations. 
The simulations are performed for various supercell sizes for flat bilayer graphene, a subset of which are reproduced below.
We use the Large-scale Atomic/Molecular Massively Parallel Simulator (LAMMPS) simulation tool that calculates the energy for a given energy functional to find the structural relaxation.
We use a reactive bond-order (REBO) intralayer potential and a registry dependent (Kolmogorov-Crespi) interlayer potential, and obtain geometry relaxed configurations using the `fire' energy minimization algorithm \cite{PLIMPTON1991,Brenner_2002,KC_2005,Ouyang2018,Bitzek_2006}.
The dislocation model requires as input material properties $C_{ijkl}$ and $A_{jl}$, that are found from energy-strain and energy-displacement simulations from atomic scale calculations.
For the classical potentials described above, we find the two independent intralayer elastic constants $C_{1111}=18.5$ eV/\AA$^2$ and $C_{1212}=5.49$ eV/\AA$^2$, and $C_{1122}=C_{1111}-2C_{1212}$.
The interface energy components are $A_{11}=A_{22}=2.52$ meV/\AA$^2$.

\begin{figure*}[h!]
\centering
\includegraphics[width=\linewidth]{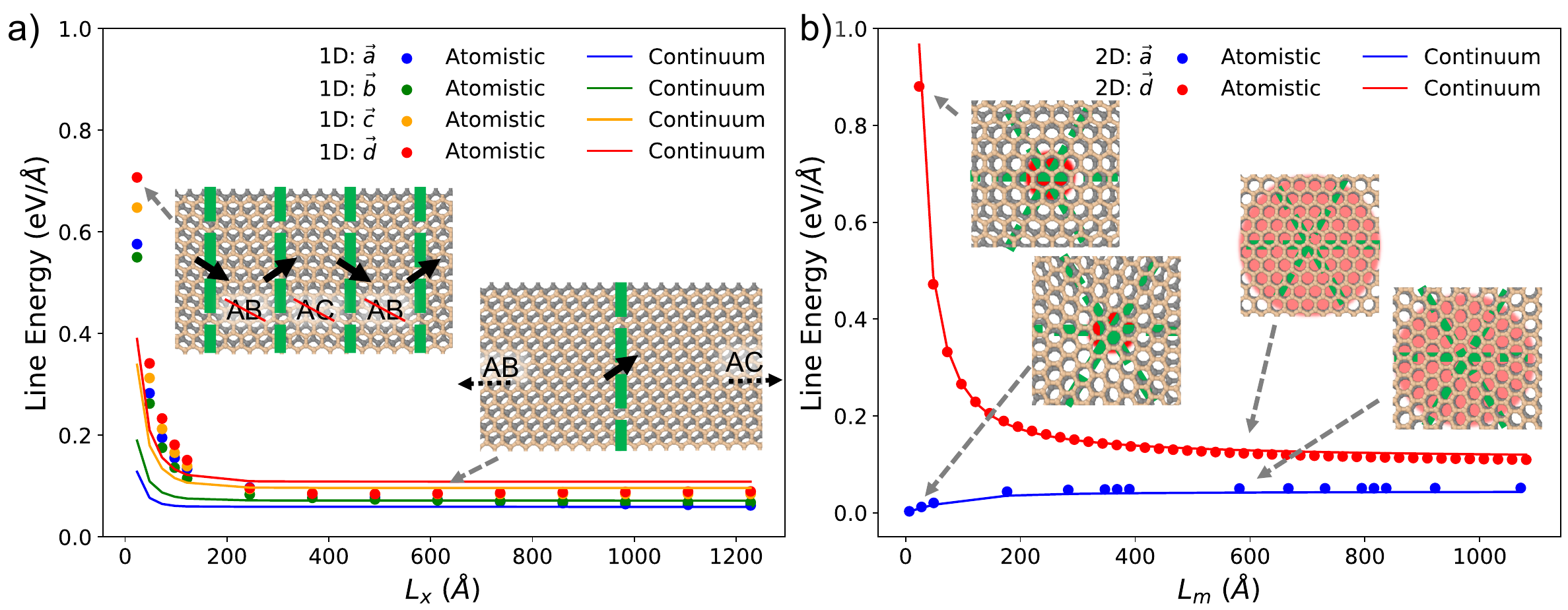}
\caption{Continuum and atomistic model for dislocation line energies for 1D and 2D dislocation networks. 
(a) The variation in the line energy with supercell length $L_x$ shows the effect of dislocation--dislocation interactions for small $L_x$ and isolated dislocations. The insets compare dislocation cores for dislocation--dislocation interactions at small supercells with a high density of dislocations and isolated dislocations with much larger cores. 
(b) Twist ($0^\circ$) and stretch ($90^\circ$)  dislocations for supercells of varying size. The insets show the dislocation junctions for small ($L_m=20$\AA) and large ($L_m=600$\AA) supercells. The uniform distortion tensor in small supercells makes the AA region much narrower than for large supercells, where the core can completely relax. It can be seen from the insets that the core regions for twist and stretch are rotated by 90$^\circ$ from each other.}
\label{fig:Evssupercell}
\end{figure*}

\begin{table*}
\centering
\begin{tabular}{ c | c c c c}
 eV/\AA& \multicolumn{4}{c}{1D Isolated Dislocations} \\
 \hline
Type & $\vec{a_i} (0^\circ)$ & $\vec{b_i} (30^\circ)$ & $\vec{c_i} (60^\circ)$ & $\vec{d_i} (90^\circ)$ \\ 
\hline
Atomistic & 0.055 & 0.065 & 0.085 & 0.093  \\
Dislocation model & 0.062 & 0.075 & 0.100 & 0.112
\end{tabular}
\caption{\label{tab:lineenergies} Dislocation energies for 1D and 2D dislocation networks normalized by the dislocation line length for large supercells ($L_x$>1000\AA).  }
\end{table*}

\subsection{Dislocation Line Energies -- 1D Networks}

The dislocation energies across supercells are reported in Figure \ref{fig:Evssupercell}.
We use the line energies -- the energy per length of dislocation -- of both small supercells with overlapping dislocation cores to large supercells with isolated dislocations.
The line energies for the four partial dislocations identified in Figure \ref{fig:1ddislocations} from 1D dislocation network supercells are shown in Figure \ref{fig:Evssupercell}(a). 
The atomistic and continuum results show the same trend, with the line energies decreasing as the dislocations become separated, converging at approximately $L_x \approx 200$ {\AA}.  
The relative line energies amongst the four partial dislocations are also in agreement, with $0^\circ$ partials having the smallest and $90^\circ$ partials having the largest line energy. 

The biggest discrepancy between the atomistic and dislocation models occurs for small supercells, where the entire supercell is out of registry due to core--core interactions. 
The left inset of Figure \ref{fig:Evssupercell}(a) shows that the high dislocation density prevents relaxation to $AB$/$AC$ stacking anywhere. 
The discrepancy is largest in this regime because the linear expansion of the interface energy in Figure \ref{fig:linearinterfacemodel}(f) is about AB/AC stacking, but since the entire supercell is everywhere far from AB/AC, the linear expansion is inadequate. 
In principle, it is possible to capture these effects by including higher order terms to better match the interface energy, but this means we could no longer solve for the distortions components separately for different $G$, since the terms would become coupled in Equation \ref{Eq:E_tot_short_flat}.

At the other extreme, isolated dislocations have cores that can completely relax (see right inset Figure \ref{fig:Evssupercell}(a)) with large regions of AB/AC stacking between them.
In this regime, the linear expansion is appropriate and the line energies from the atomistic and dislocation models for each dislocation agree well.
The plateau of the dislocation line energy for large $L_x$ means that the dislocations are indeed isolated as there are no long-range strain fields interacting.
The line energies of the isolated dislocations are produced in Table \ref{tab:lineenergies} to show the quantitative agreement.

\subsection{Dislocation Line Energies -- 2D Networks}

The line energies of twist and stretch moir\'e patterns are compared across supercell sizes in Figure \ref{fig:Evssupercell}(b).
The line energies of the 2D  $0^\circ$ and $90^\circ$ dislocation networks have nearly identical trends for the atomistic and dislocation descriptions.
Notably, the shape of the line energies across supercells for $\vec{a}$ and $\vec{d}$ dislocation networks are different, but the dislocation model accurately reproduces the opposing trends.
The change of shape is due to the dislocation junctions present in 2D networks, whose energy is constant and negative (positive) for $0^\circ$ ($90^\circ$) dislocations respectively.

In contrast to 1D dislocation networks, the line energy for 2D dislocation networks is in good agreement for both supercells with dislocation-dislocation interactions (small $L_m$) and isolated dislocations (large $L_m$).
The good agreement for large supercells is expected, since as for the 1D case the interfacial energy in the large regions of AB stacking in the interior are well described in our model. 
The good agreement for the smaller supercells is more surprising, but occurs directly as a result of the topological constraint imposed by the 2D dislocation network. 
This constraint forces the interior of the triangular regions to have AB/AC stacking, no matter what the size of the moir\'e superlattice. 
The effect of the supercell size on the absolute size of the AA stacking region is seen in the insets of Figure \ref{fig:Evssupercell}(b).
The pair of insets corresponding to small $L_m$ show a small AA region (red), and by necessity maintain AB/AC stacking between the junctions. 
This is true even though the relative proportion of AA stacking present in the supercell is larger for small supercells.

\begin{figure}[t]
\centering
\includegraphics[width=0.5\linewidth]{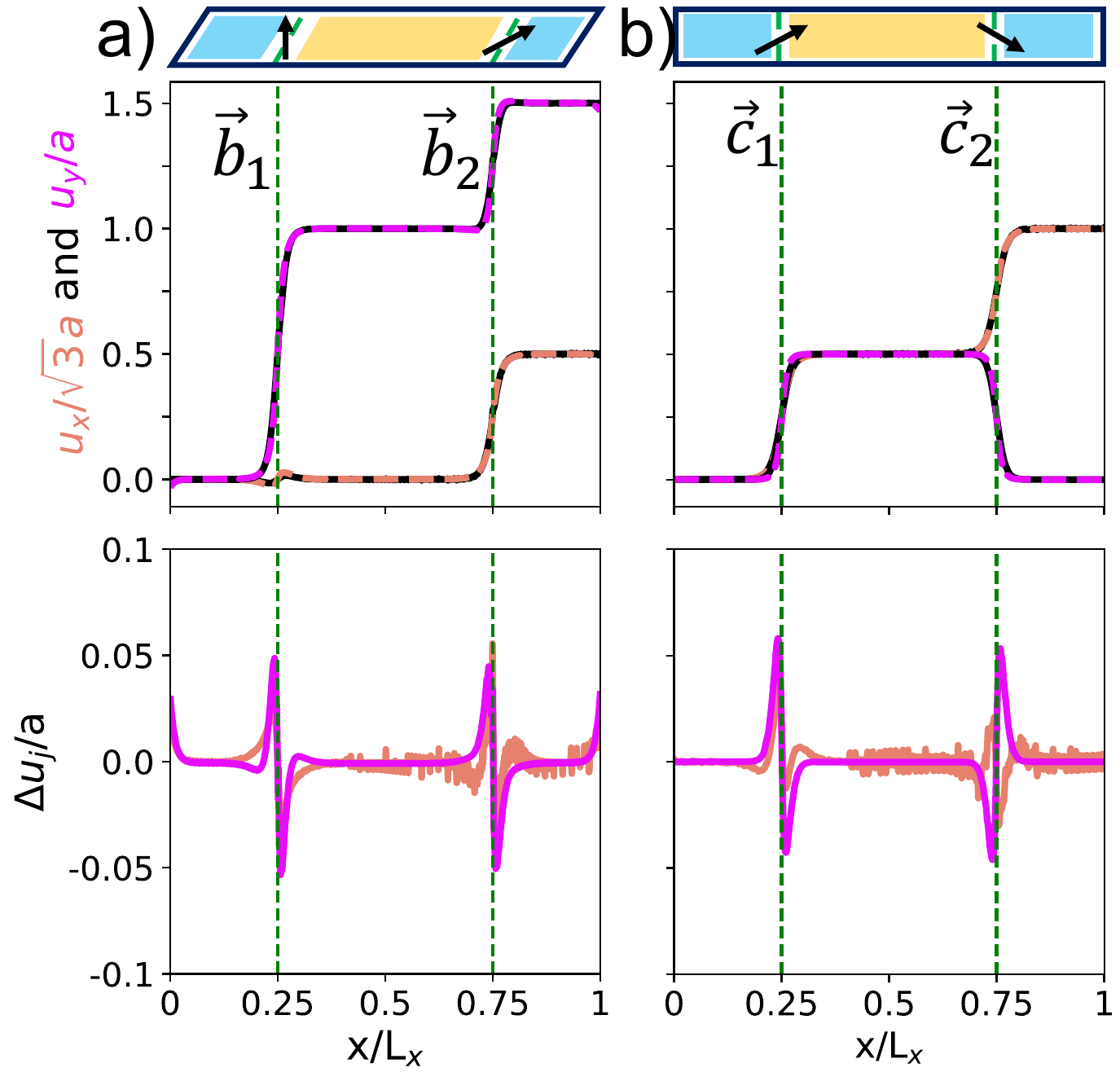}
\caption{Atomistic (solid, black) and continuum (dotted) description of displacement fields for 1D partial dislocation networks for supercell with $L_x = $2500 \AA.
Salmon/magenta represent displacements in the $x$/$y$ direction, for (a) two 30$^\circ$ dislocations and (b) two 60$^\circ$ dislocations.
The bottom row shows the deviation between the atomistic and continuum displacement fields. }
\label{fig:1ddisplacement}
\end{figure}

\subsection{Structural Relaxations -- 1D Networks}

The structural relaxations for 1D networks from the atomistic and dislocation simulations are compared for supercells with $L_x =$ 2500 \AA.
The displacement fields for each simulation are normalized by the carbon-carbon spacing {\it a} or lattice spacing $\sqrt{3}a$ to highlight the symmetries of the dislocations.
The displacement fields of networks of 1D partial dislocations in Figure \ref{fig:1ddisplacement} show agreement between continuum and atomistic for $\vec{b}_i$ and $\vec{c}_i$ partial dislocations; $\vec{a}_i$ and $\vec{d}_i$ partials are presented in the Supplementary Information (and will appear here later in the 2D networks).
The displacement fields are shown in the top row, where the $x$ and $y$ components of the dislocation model are shown in dashed salmon and magenta respectively and the atomic simulation is shown in black.
The dislocation model reproduces the atomic simulation, where the solid black line is nearly obscured by the dislocation model.
Impressively, the dislocation model picks up small features of the atomistic results at the dislocation core (u$_x$ at x/L$_x$ = 0.25 for Figure \ref{fig:1ddisplacement}(a)).
The deviation between atomistic and continuum displacement fields is shown in the bottom row.
The normalized difference shows a maximum difference of 6\%, less than 0.1 Å.

\begin{figure*}[t]
\centering
\includegraphics[width=6.45in]{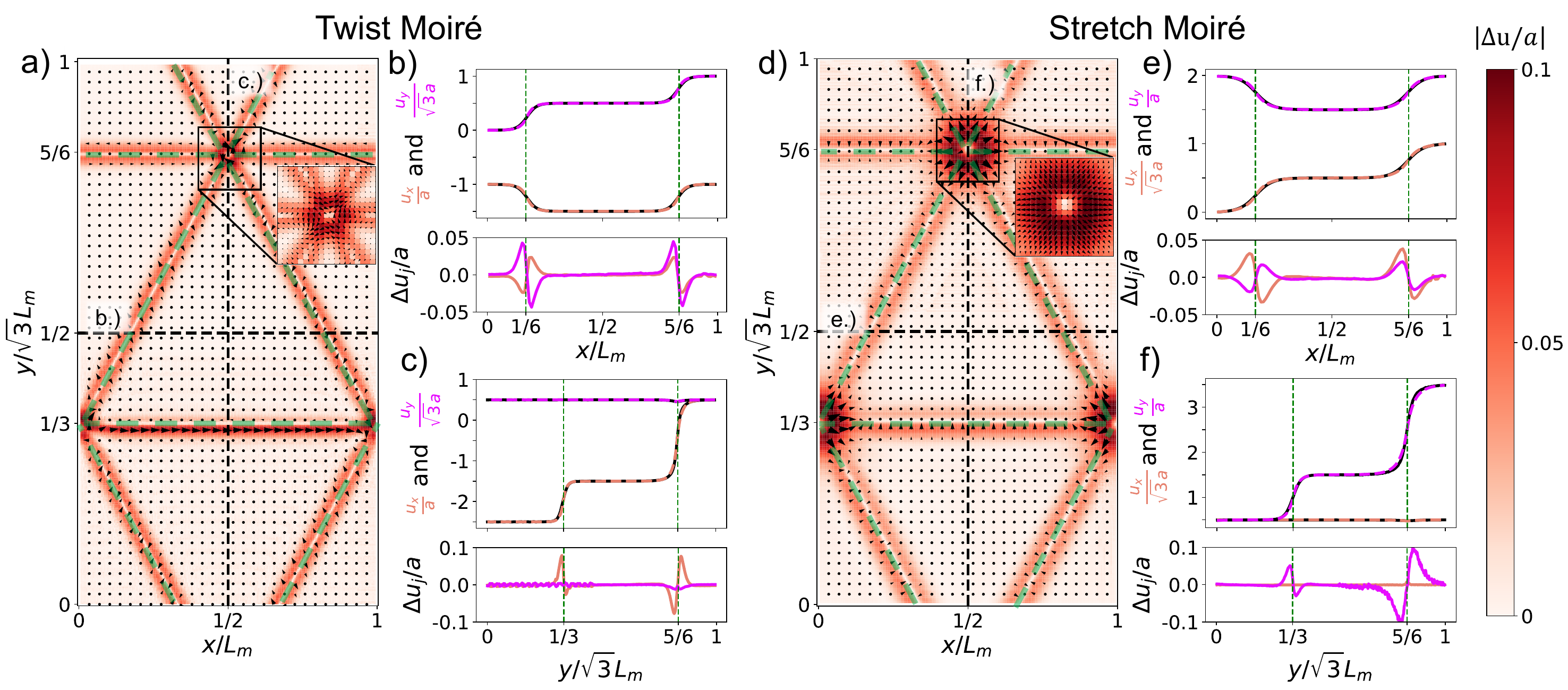}
\caption{Atomistic and continuum displacement fields for 2D partial dislocation networks for (a-c) twist moir\'e and (d-f) stretch moir\'e patterns.
(a) Deviation between atomistic and continuum  ($\Delta u_j=u_j^{cp}-u_j^{dm}$) for a twist angle of  $\theta = 0.13^\circ$ ($\lambda_m$ = 107 nm) moir\'e pattern.
The contour plot shows the magnitude of difference, while the vector field shows the direction.
(b,c) Line scans in the moir\'e zig-zag/armchair direction showing the classical potential (solid) and continuum model (dotted), $x$ (blue) and $y$ (orange) displacement fields and their difference.
(d) Difference between classical potential and dislocation model displacement fields for a stretch $\varepsilon$ = 0.1\% ($\lambda_m$ = 108 nm) moir\'e pattern.
(e,f) Line scans in the moir\'e zig-zag/armchair direction showing the classical potential (solid) and continuum (dotted) displacement fields and their difference. 
Insets in (a,d) show the different structures of 0$^\circ$ and 90$^\circ$ partial dislocation junctions.}
\label{fig:2ddisplacement}
\end{figure*}

\subsection{Structural Relaxations -- 2D Networks}

The structures predicted by the atomistic and continuum approach for 2D dislocation networks are compared in Figure \ref{fig:2ddisplacement}. 
The deviation between the two approaches is plotted by the contour plot on the rectangular unit cells for both twist and stretch moir\'e superlattices for $L_m = 1080$ {\AA}.
A quiver plot that shows the direction and magnitude of the difference is overlaid.
In both the twist and stretch moir\'e superlattices, there is good agreement with  maximum errors of 10\% localized to the dislocation junctions and of 5\% at the dislocation lines.
Insets show the dislocation junctions in greater detail, which show the different reconstructions present for twist and stretch junctions.

Two line scans of the displacement fields are shown in Figure \ref{fig:2ddisplacement}.
A horizontal line scan taken at $y/\sqrt{3}$L$_m$=1/2 crosses $\vec{a}_i$, $\vec{d}_i$ dislocations at $x/L_m$=1/6 and 5/6 and shown in Figure \ref{fig:2ddisplacement}(b,e) respectively. 
The twist moir\'e line scan shows that the two dislocations have opposite $x$ components but the same $y$ component.
The two dislocations in the stretch moir\'e line scan, however, have opposite $y$ components but the same $x$ component showing how the two superlattices differ by a 90$^\circ$ rotation.
The horizontal line scans crossing isolated dislocations show good agreement between the displacement fields with less than 5\% normalized error at any location.

A second vertical line scan of twist and stretch superlattices is taken at $x/L_m$=1/2 is shown in Figure \ref{fig:2ddisplacement}(c,f).
The line scans cross a single dislocation perpendicularly at $y/\sqrt{3}L_m=1/3$ and show that the Burgers vector for twist (stretch) moir\'e patterns are parallel (perperpendicular) to the dislocation line and have a magnitude of a.
The line scans cross a dislocation junction at $y/\sqrt{3}$L$_m$=5/6 revealing that junctions have twice the Burgers vector of a single dislocation.
The difference of the displacement fields shows that the maximum normalized error is just less than 10\% at the dislocation junctions.

\section{Applications of Continuum Dislocation Framework}

Having established the energy and structural correspondence between the continuum dislocation model and results of atomistic simulations, we now highlight some possible applications of the model.

\subsection{Structural Trends of Moir\'e Superlattices}

\begin{figure*}[t]
\centering
\includegraphics[width=\linewidth]{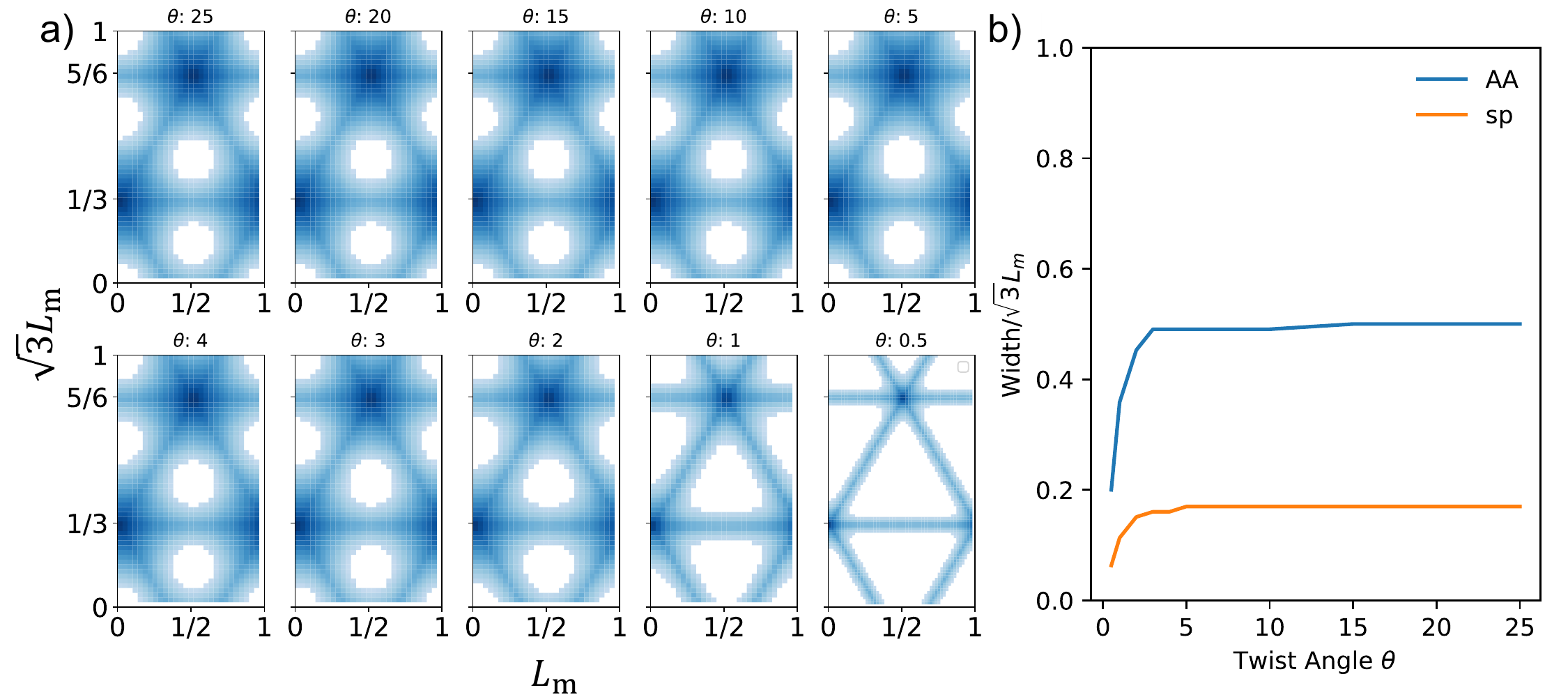}
\caption{Structure of twist moir\'e superlattices versus twist angle. 
(a) Heat maps of the disregistry across twist angle $\theta$, where the blue indicates disregistry $> a/4$, or half of the maximum value. 
(b) The width of AA and SP stacking regions across twist angle $\theta$.}
\label{fig:structurevstheta}
\end{figure*}

Previous studies have demonstrated that the relative size of the AA stacking regions grow with increasing twist angle \cite{ZHANG2017,Gargiulo_2017}. 
Based on our theory, this effect can be understood to arise from the necessity to satisfy the topological constraints of the dislocation network
even as the decreasing superlattice size (increased twist) confines the network. 
The blue color maps in Figure \ref{fig:structurevstheta}(a) show the regions of large disregistry, defined here as $r_m > a/4$. for varying twist angle $\theta$. 
For large twist angle ($\theta > 2^\circ$), the portion of the superlattice unit cell exhibiting $r_m > a/4$ is similar and relatively large. 
As $\theta$ decreases below $2^\circ$, both the junctions and the dislocation lines themselves take up a smaller proportion of the superlattice area and the large triangular regions of AB/AC stacking emerge. 
The proportion of dislocated regions across twist angles is compared quantitatively in Figure \ref{fig:structurevstheta}(b) for both SP and AA stacking.
It confirms the visual analysis from Figure \ref{fig:structurevstheta}(a), the relative size of the dislocation regions is similar for twist angle $\theta>2^\circ$, but decreases for smaller twist angles, where the dislocations are fully relaxed due to large supercell size $L_m$.

\begin{figure}[t]
\centering
\includegraphics[width=0.5\linewidth]{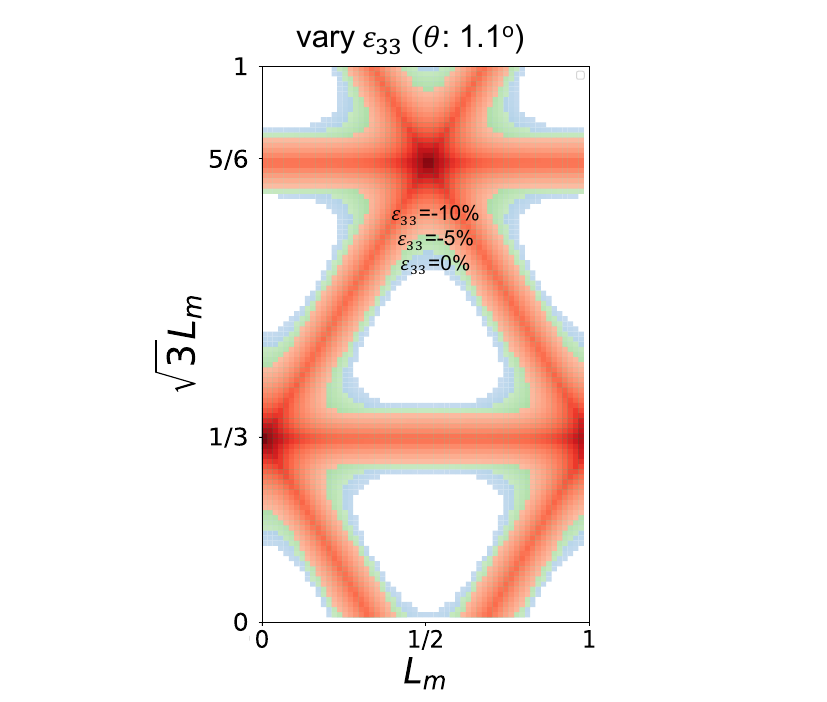}
\caption{Structure of twist moir\'e patterns for varying compressive strain $\varepsilon_{33}$ for constant $\theta$.
Color maps represent different $\varepsilon_{33}$, where the color represents deviations from AB/AC stacking greater than $a/4$. 
Magic-angle twisted bilayer graphene ($\theta$ =  1.1$^\circ$) at equilibrium interlayer spacing is used as a reference (blue). 
Green represents a compression of 5\% ($d_z$: 3.23\AA), while red represents a 10\% compression ($d_z$: 3.06\AA).}
\label{fig:varalphatheta}
\end{figure}

In addition, our model can address structural relaxation due to out--of--plane compression by refitting the interface energy parameter $A_{jl}$ for different interlayer spacing (details of the fitting are in the SI).
We examine how compressing the bilayers in the out--of--plane direction can affect the moire structure for a given twist angle. 
Compressing the bilayers this way has been shown to tune the 'magic' angle in bilayer graphene \cite{Yankowitzeaav2019}.
In Figure \ref{fig:varalphatheta}, blue, green, and red correspond to compression with $\epsilon_{33}$= 0\%, -5\%, and -10\% for the `magic' twist angle of $\theta = 1.1^\circ$\cite{Bistritzer2011}.
The trends shown may indicate how compression can tune the magic angle by modifying the structure, since increased compression reduces the relative size of the dislocation cores and junctions (similar to the effect of reducing the twist angle).

\begin{figure}[t]
\centering
\includegraphics[width=0.5\linewidth]{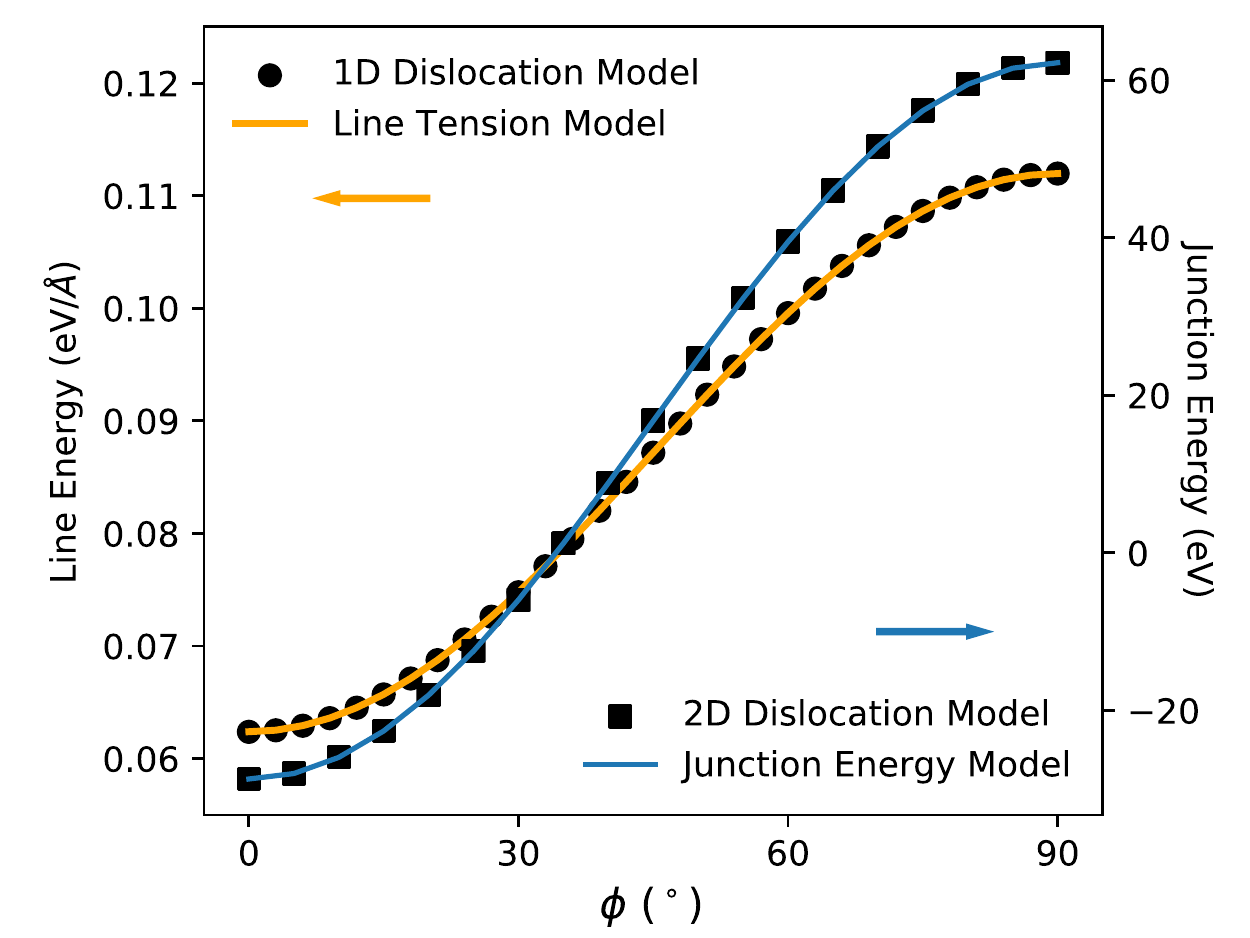}
\caption{
Line tension (yellow) of a single dislocation and dislocation junction energy (blue) as a function of  angle $\phi$. 
The energies can be approximated within a line tension model that uses only two parameters. 
The negative junction energy for small $\theta$ suggests favorable dislocation interactions and dislocation--dislocation attraction. }
\label{fig:linetensionmodel}
\end{figure}

\subsection{Dislocation line and junction models}

Finally, we use the dislocation model to estimate the dislocation line and junction energies for arbitrary $\phi$. 
These quantities could be used to drive meso-scale dislocation dynamics simulations to explore, for instance, how the moir\'e structure interacts with external strain fields \cite{KUBIN2001}.
We investigate both the dislocation line energies and the dislocation junction energies.

Figure \ref{fig:linetensionmodel} shows the line energies for continuous $\phi$.  
These energies are obtained from the continuum formalism, and compared to the approximate functional form 
\begin{equation}
E_{l}(\phi) = E_{l}(90^\circ)-(E_{l}(90^\circ)-E_{l}(0^\circ))\cos^2(\phi) \hspace{0.2em},
\label{Eq:lineenergy}
\end{equation}
where $E_l(\phi)$ is the dislocation line energy, and $E_{l}(0^\circ)$, $E_{l}(90^\circ)$ are obtained from Table \ref{tab:lineenergies}, and the factor $\cos^2(\phi)$ comes from the line tension approximation of 3D dislocations\cite{cai_nix_2016}. 
The two quantities show a good correspondence. 

Meanwhile to estimate junction energies, we calculate the energy of 2D dislocation networks with well separated cores ($L_m$=500Å) and subtract the energy associated with the dislocation lines from Eq. \ref{Eq:lineenergy}.
The remaining energy is the junction energy. 
Figure \ref{fig:linetensionmodel} shows that the dislocation junction energies are not uniform with $\phi$. 
Instead, the 0$^\circ$ dislocation junctions have negative energy while 90$^\circ$ junctions have positive energies with a crossover around 34$^\circ$.
This finding is consistent with Figure \ref{fig:Evssupercell}(b), which showed opposite trends for $0^\circ$ and $90^\circ$ with decreasing $L_m$. 
The same functional form from Eq.~\ref{Eq:lineenergy} is used fit to the dislocation energy, using junction energy $E_j$ rather than line energy $E_l$ using $E_{j}(0^\circ)$= -28.7 eV and $E_{j}(90^\circ)$ = 62.2 eV as boundary conditions.

The energy landscape of the dislocation line and junction energies reveal that 2D dislocation networks favor 0$^\circ$ dislocations.
This may be the origin of the non-uniform moir\'e superlattices observed experimentally, for instance in dark-field transmission electron microscopy images of Alden {\it et al}.  \cite{Alden2013}
Instead of a uniform moir\'e period over microns, the dislocation networks relax to maximize the amount of 0$^\circ$ dislocations and junctions.

\section{Conclusion}

We have presented a dislocation theory based on topological constraints to describe interlayer dislocations in bilayer graphene. 
In our approach, both 1D and 2D (moir\'e) superlattices are defined in terms of the periodic dislocation networks of which they are comprised. 
Conventional dislocation theory is adapted so as to treat the discrete nature of each layer of the 2D bilayer by describing the total energy as arising from both the elastic energy of each distorted layer, together with an interface energy that couples the layers.
The dislocation model does not assume any analytic form for the solution, naturally accounts for dislocation-dislocation interactions, and contains no adjustable parameters.
The energy and structure predictions of the dislocation model are in agreement with atomic scale calculations.
Finally, we present two applications of our model: an investigation of the evolution of the atomic scale structure as a function of moir\'e twist angle, and prediction of line tension and dislocation junction energies for arbitrary dislocation sense $\phi$.

\section{Acknowledgements}

We gratefully acknowledge the grants that supported this research.  EA and HJ acknowledge the support of the Army Research Office (W911NF-17-1-0544) Material Science Division under Dr. Chakrapani Varanasi.  The partial support of NSF grant number CMMI 18-25300 (MOMS program) is also acknowledged.  
In addition, EA and EE acknowledge the support of the National Science Foundation through the Illinois Materials Research Science and Engineering Center (I-MRSEC) under Grants No. DMR-1555278 and DMR-1720633.  
Helpful conversations with Dr. Pascal Pochet, Dr. Josh Schiller, Dr. Jaehyung Yu, and Tawfiq Rakib are also gratefully acknowledged.

\bibliographystyle{ieeetr}
\bibliography{references}
\end{document}